\documentclass[%
reprint,
nofootinbib,
 amsmath,amssymb,
 aps,
 prb,
]{revtex4-2}

\usepackage{graphicx}
\usepackage{dcolumn}
\usepackage{bm}
\usepackage{xcolor}

\begin{document}


\title{Doping dependence of the upper critical field in untwinned YBa$_2$Cu$_3$O$_{7-\delta}$ thin films}

\author{Eric Wahlberg}
\email{ericand@chalmers.se}
\affiliation{Quantum Device Physics Laboratory, Department of Microtechnology and Nanoscience, Chalmers University of Technology, SE-41296, Göteborg, Sweden}
\author{Riccardo Arpaia}
\affiliation{Quantum Device Physics Laboratory, Department of Microtechnology and Nanoscience, Chalmers University of Technology, SE-41296, Göteborg, Sweden}
\author{Alexei Kalaboukhov}
\affiliation{Quantum Device Physics Laboratory, Department of Microtechnology and Nanoscience, Chalmers University of Technology, SE-41296, Göteborg, Sweden}
\author{Thilo Bauch}
\affiliation{Quantum Device Physics Laboratory, Department of Microtechnology and Nanoscience, Chalmers University of Technology, SE-41296, Göteborg, Sweden}
\author{Floriana Lombardi}%
 \email{floriana.lombardi@chalmers.se}
\affiliation{Quantum Device Physics Laboratory, Department of Microtechnology and Nanoscience, Chalmers University of Technology, SE-41296, Göteborg, Sweden}%

\date{\today}
\begin{abstract}
We report on measurements of the doping dependence of the upper critical field $H_{c,2}$ in 50 nm thick YBa$_2$Cu$_3$O$_{7-\delta}$ films. The films are untwinned and are characterized by a small in-plane compressive strain. We find that the $H_{c,2}$ shows a strong decrease in the underdoped region of the phase diagram, in agreement with what has been measured in relaxed single crystals. The origin of the decrease of $H_{c,2}$ in the underdoped regime is discussed within a scenario where charge density wave order competes with superconductivity. This demonstrates the potential of using thin films for studying the phase diagram of high-$T_c$ materials under strain.

\end{abstract}

\maketitle

\section{Introduction}
The origin of superconductivity in high-$T_c$ cuprates is still an open question, mainly due to the complex intertwining of various electronic orders with the superconductive phenomenon itself. In the recent years the relation between charge density wave (CDW) order and superconductivity has been the subject of intensive investigation \cite{Ghiringhelli2012,Chang2012,Blanco2014,Cyrchoinere2018,arpaia2021charge}. It is by now clear that the two orders are competing, which is supported by the strong suppression of $T_c$ at the hole-doping level of $p\approx0.12$, where CDW order is strongest \cite{Chang2012,Ando2004,Liang2006,huecker2014competing}. Another sign of the competition is the rapid decrease of the upper critical field $H_{c,2}$ with decreasing doping found in YBa$_2$Cu$_3$O$_{7-\delta}$ (YBCO) single crystals \cite{ando2002magnetoresistance,ramshaw2012vortex,chang2012decrease,grissonnanche2014direct}. According to the thermodynamic Ginzburg-Landau theory a measure of $H_{c,2}=\Phi_0/2\pi\xi^2$, where $\Phi_0$ is the flux quantum and $\xi$ the superconducting coherence length, allows to retrieve the doping dependence of the coherence length \cite{ando2002magnetoresistance}.

Recent experiments have shown that in YBCO $H_{c,2}(p)$ has two maxima: one in the underdoped regime ($p\approx0.08$) and one in the overdoped regime ($p\approx0.18$) \cite{grissonnanche2014direct}. These doping levels do not correspond to the maximum superconducting critical temperature $T_c$, but appear to be connected to the extremes of the doping range where CDW order is present in the phase diagram \cite{Blanco2014}, indicating a strict relation between the two. However, the measurement of $H_{c,2}$ as a function of doping in the cuprates has been controversial \cite{ando1999resistive}, with some experiments pointing toward an increase (rather than a decrease) with decreasing doping \cite{wang2003dependence}.  

To better understand which factors determine the doping dependence of $H_{c,2}$ one
can study how it is affected by a tuning parameter, such as strain. Recent papers have shown that the CDW order can be tuned by strain externally applied to single crystals \cite{kim2018uniaxial,kim2021charge,Cyrchoinere2018,Souliou2018, choi2022unveiling}. They have found that $T_c$ increases and the CDW order is strongly modified \cite{Cyrchoinere2018}. In thin films the strain can be tuned by the choice of substrate and by varying the film thickness \cite{arpaia2019,wahlberg2021}, contrary to the case of single crystals where complicated apparatus are required to apply the strain which might not be compatible with high magnetic field facilities and/or other measurement setups.

Here we show the growth of high-quality thin films of YBCO covering a large part of the phase diagram from $p=0.10$ to $p=0.18$ \cite{arpaia2018,baghdadi2014toward}. We have previously shown that For 50 nm thick slightly overdoped films ($p=0.18$) we can generate a slight compressive uniaxial in-plane strain by using surface reconstructed MgO(110) substrates. These films are also untwinned \cite{arpaia2019}. Here we present structural and electrical transport characterization of 50 nm thick YBCO films on MgO(110) substrates as a function of the doping and show that the underdoped films are untwinned and are strained similarly to the slightly overdoped films. From measurements of the superconducting resistive transition of the films as a function of an applied magnetic field we extract the doping dependence of $H_{c,2}$. The results show that the $H_{c,2}(p)$ behavior in our thin films is in agreement to what has been measured in single crystals \cite{grissonnanche2014direct}, with a strong decrease in the underdoped regime. We discuss the origin of this similarity within a scenario where CDW order and superconductivity compete.

\begin{figure}[t]
\centering
\includegraphics[width=0.3\textwidth]{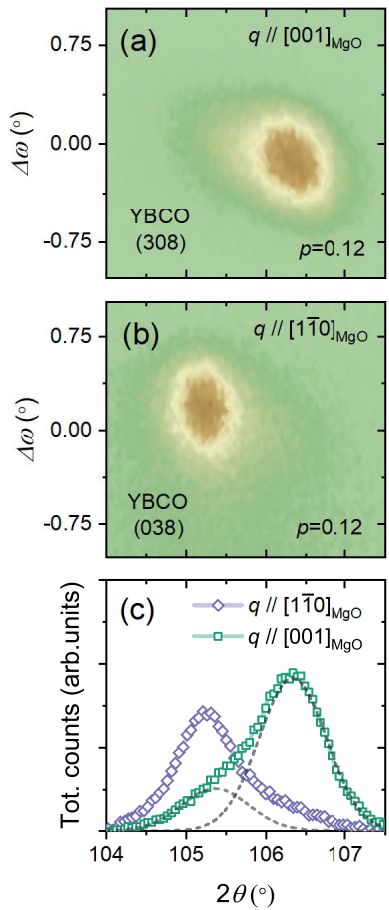}
\caption{\label{fig:XRD} XRD structural characterization of an underdoped YBCO thin film to determine the in-plane structure.(a,b) 2$\theta$-$\omega$/$\Delta\omega$ intensity maps around the YBCO (308) and (038) Bragg reflections measured along the two orthogonal substrate directions MgO [1$\overline1$0] and [001] in an underdoped YBCO thin film ($p$=0.12). (c) Integrated intensity versus 2$\theta$ from the maps in (a) and (b). The gray dashed lines shows Gaussian fits of the two peaks that build up the total intensity. }
\end{figure}

\section{Untwinned and underdoped YBCO thin films}

\begin{figure}[t]
\centering
\includegraphics[width=0.4\textwidth]{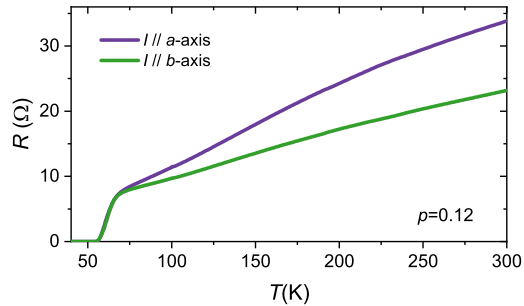}
\caption{\label{fig:Res} Temperature dependence of the resistance measured in an untwinned, underdoped YBCO thin film with the current $I$ applied along the YBCO $a$- and $b$-axis (purple and green line).
} 
\end{figure}

The YBCO films used in this experiment are 50 nm thick and are grown by pulsed laser deposition on (110) oriented MgO substrates following a procedure described elsewhere \cite{arpaia2018,arpaia2019}. The films span a wide range of hole-doping $p$, going from underdoped ($p\approx0.10$) up to slightly overdoped ($p\approx0.18$). In this range of $p$ the CDW order grows in strength as the doping is reduced \cite{Blanco2014} which allows to study the possible competition between CDW and superconducting orders. To promote the growth of untwinned YBCO films the substrates are annealed at high temperature ($T=790 ^\circ$C) before the film deposition to allow surface reconstruction \cite{arpaia2019}. The untwinning ratio is estimated from a characterization of the crystal lattice of the YBCO films. Figure \ref{fig:XRD}(a,b) shows X-ray diffraction (XRD) asymmetric $2\theta$-$\omega/\omega$ intensity maps around the (308) and (038) YBCO Bragg reflections for a film at $p \approx 0.12$. The two reflections are well separated in 2$\theta$, a consequence of the orthorhombicity induced by the one-dimensional CuO chains oriented along the YBCO $b$-axis \cite{Jorgensen1990}. We observe that there is a minor (038) component in the (308) map and vice versa, as visualized by the integrated intensity plot in Figure \ref{fig:XRD}(c). By the ratio of these two components (see gray dashed lines) we estimate that the films are 83\% untwinned. From the $2\theta$ values of the (308) and (038) Bragg reflections and the (00L) Bragg reflections (not shown here) we have extracted the values of the in-plane lattice parameters $a$ and $b$ and the out-of-plane lattice parameter $c$ (see table \ref{tab}). 
Compared to YBCO crystals of similar doping \cite{Jorgensen1990} the $b$-axis is slightly shorter and the $c$-axis slightly longer in the thin films, which is indicative of a uniaxial in-plane compressive strain. The level of in-plane strain ($\delta b =-0.5\%$) is close to what we have previously reported on slightly overdoped YBCO thin films on MgO (see table \ref{tab}) which means that the strain induced by the substrate does not change much with the doping level of the film.
The temperature dependence of the electrical resistance along the YBCO $a$- and $b$-axis of an untwinned film measured in a four-point Van der Pauw configuration is shown in figure \ref{fig:Res}. The resistivity anisotropy (between the YBCO $a$- and $b$-axis) is slightly lower than that of completely untwinned crystals at this doping \cite{Ando2004} which is consistent with the 83\% untwinning ratio of our films. CuO chains along the $b$-axis in YBCO are the origin of the orthorhombicity of the unit cell. In electrical transport they cause an in-plane resistivity anisotropy as they are weakly conducting \cite{gagnon1994t,Ando2002}. 

\begin{table}[t]
\centering
\begin{tabular}{ |l|c|c|c| }
 \hline
 $p\approx$ 0.12 (0.18) & $\textbf{a}$ & $\textbf{b}$ & $\textbf{c}$ \\ 
 \hline 
 \textbf{Crystal} & 3.83 (3.82)& 3.89 (3.89)& 11.72 (11.69)\\ 
 \hline
 \textbf{Thin Film} & 3.83 (3.82)& 3.87 (3.87)& 11.73 (11.71)\\ 
 \hline
\end{tabular}
\caption{\label{tab} Comparison of lattice parameters (in units of Å) of underdoped (slightly overdoped in parenthesis) YBCO thin films and crystals \cite{Jorgensen1990,arpaia2019}. 
} 
\end{table} 

\begin{figure*}[t]
\centering
\includegraphics[width=0.8\textwidth]{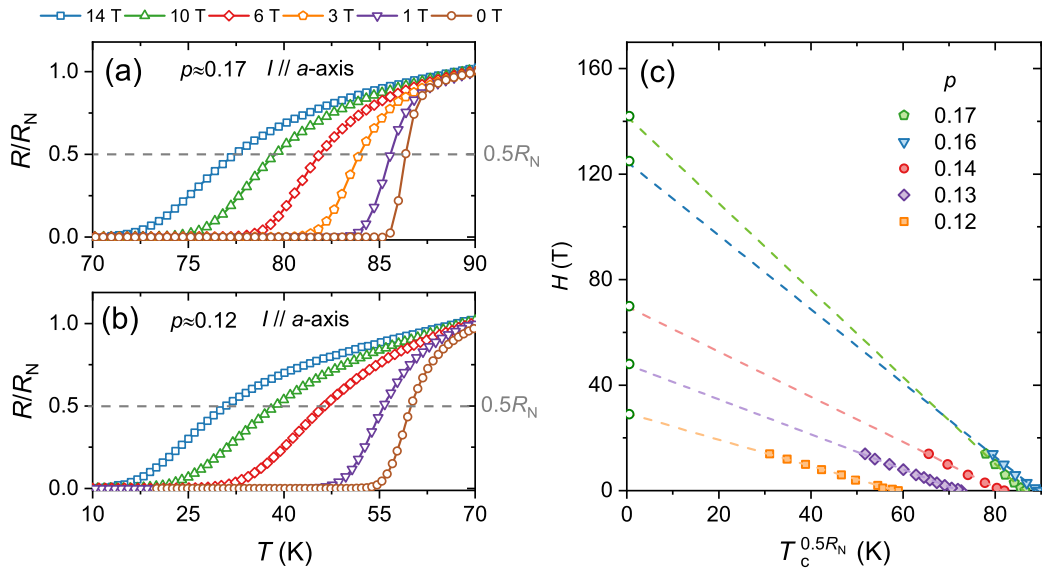}
\caption{\label{fig:Field}(a,b) Magnetic field dependence of the resistive transition in YBCO films of various doping ((a) $p=0.17$, (b) $p=0.12$). The magnetic field is applied along the $c$-axis, perpendicular to the current which is applied along the $a$-axis. The gray dashed lines indicate where the resistance has dropped to 50\% of normal state resistance $R_N$ at $T_c$. (c) Linear fits to $T_c^{0.5R_N}(H)$ for films of different values of $p$. The green circles show the linear extrapolated values of $H_{c,2}$ at $T_c^{0.5R_N}=0$ 
} 
\end{figure*}

\section{Estimation of $H_{c,2}$}
Measurements of the temperature dependence of the electrical resistance $R(T)$ in a range of temperature around $T_c$ have been performed for 50 nm thick YBCO thin films as a function of doping and magnetic field. Figure \ref{fig:Field}(a,b) shows the magnetic field dependence of the resistive transition in a slightly overdoped ($p\approx0.17$) and an underdoped ($p\approx0.12$) film. The magnetic field $H$, spanning the range from 0 to 14 T, is applied perpendicular to the film surface. For each film we have extracted the magnetic field dependence of the critical temperature $T_c^{0.5R_N}$, which is defined as the temperature where the resistance has dropped to 50\% of normal state value $R_N$ (see gray dashed line in Figure \ref{fig:Field}(a,b)), where $R_N$ is the normal state resistance above the superconducting transition. $H_{c,2}$ is obtained by making a linear fit to $T_c^{0.5R_N}(H)$ and extrapolating to the zero temperature value \cite{ando1999resistive}. Figure \ref{fig:Field}(c) shows $T_c^{0.5R_N}(H)$ and the linear fits for the YBCO thin films of different doping. We observe a clear trend: in the films with lower $p$ the magnetic field has a stronger effect on the superconducting transition, indicating a lower $H_{c,2}$ value.

\begin{figure}[b]
\centering
\includegraphics[width=0.49\textwidth]{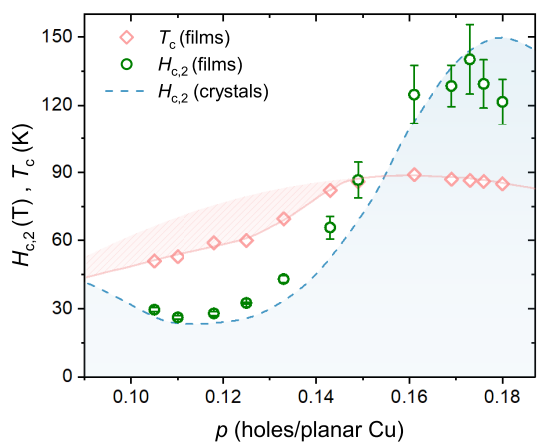}
\caption{\label{fig:Hc2} Doping dependence of $H_{c,2}$ (green circles) and $T_c$ (pink squares) for the YBCO thin films. The blue dashed line shows the doping dependence of  $H_{c,2}$ measured in YBCO crystals \cite{grissonnanche2014direct}. The pink shaded area shows where $T_c$ is suppressed below the parabolic doping dependence \cite{arpaia2018}. 
}
\end{figure}

The doping dependence of $H_{c,2}$ extracted in figure \ref{fig:Field}(c) is presented in figure \ref{fig:Hc2}. The values reported come from measurements along the YBCO $a$-axis, but there is no significant difference in extracting the values from measurements along the $b$-axis (data not shown here). We find that $H_{c,2}$ drops quickly from $\approx$ 140 T to $\approx$ 30 T when the doping is reduced from slightly overdoped to underdoped. This doping trend of $H_{c,2}$ is in excellent agreement with what has been observed in single crystals by various measurement methods (see blue dashed line in figure \ref{fig:Hc2}) in the underdoped regime \cite{grissonnanche2014direct,ramshaw2012vortex,ando2002magnetoresistance}. We therefore conclude that the small in-plane compressive strain induced by the MgO substrate does not have any significant effect on the critical field. The mechanism that causes the drop of $H_{c,2}$ with doping is not completely settled, although it is likely associated to the appearance of CDW order. As mentioned previously, CDW order is present in YBCO in the doping range $p\approx$ 0.08-0.17 \cite{Blanco2014}. In the same doping range there is a suppression of $T_c$ compared to the expected approximately parabolic $p$ dependence \cite{Ando2004,Liang2006}, which is also associated to CDW order \cite{huecker2014competing}. We have previously reported that in similar 30 nm thick YBCO films the suppression of $T_c$ around $p=0.125$ is reproduced (see pink shaded area in figure \ref{fig:Hc2}) \cite{arpaia2018, Andersson2020}. Moreover, we have found by RIXS that CDW order is present in our 50 nm thick films, with characteristics close to those found in single crystals \cite{wahlberg2021}. If the drop of $H_{c,2}$ is a consequence of CDW order, it is therefore not surprising that we find similar values in our thin films as in crystals. It remains to be seen why the uniaxial compressive strain in our film does not produce the same effect on the CDW order as instead happens in single crystals \cite{kim2018uniaxial}.

In the slightly overdoped regime we find that our data deviates from those reported on single crystals. In the crystal measurements the maximum $H_{c,2}$ has been estimated to be close to $p\approx0.18$, though there is a significant uncertainty in this estimation \cite{grissonnanche2014direct}. The exact doping level of the maximum $H_{c,2}$ is crucial to understand the origin of the reduction in the underdoped regime. Some theories of high-$T_c$ superconductivity consider that superconductivity stems from the quantum fluctuations associated to a quantum critical point (QCP) \cite{Keimer2015}. A QCP close to optimal doping in YBCO has been postulated as the end point of the 3D CDW order around $p=0.16$ corresponding to optimal doping \cite{castellani1996non, arpaia2019dynamical}. Alternatively, or in addition to that, a putative QCP has been associated to the end point of the enigmatic pseudogap phase at $p\approx$ 0.19 \cite{ramshaw2015quasiparticle,michon2019thermodynamic}. The maximum in $H_{c,2}(p)$ in our measurements is closer to $p\approx$ 0.17, which tends to exclude the pseudogap as the origin of the suppression of $H_{c,2}$ at low doping. Nevertheless, the decreased value of $H_{c,2}$ below $p\approx$ 0.17 could be related to the reduced number of states available to superconductivity due to the pseudogap opening and due to the reconstruction of the Fermi surface induced by CDW \cite{leboeuf2007electron}, as some experiments seems to support.

A puzzling observation is that the maximum of $H_{c,2}$ at $p\approx 0.17$ does not correspond the the maximum of $T_c$, which is at $p=0.16$ (see figure \ref{fig:Hc2}) indicating that the correspondence between the upper critical field and superconductivity is more complex than what is described by the phenomenological Ginzburg-Landau theory. 

\section{Conclusions}
We have presented measurements of the upper critical field, $H_{c,2}$ as a function of the doping in 50 nm thick YBCO films. The films are untwinned with a small in-plane compressive strain along the $b$-axis. $H_{c,2}(p)$ was estimated by characterizing the magnetic field dependence of the resistive transition. We found that the doping dependence of $H_{c,2}$ measured in single crystals is reproduced in our thin films, with the exception of the doping level above the optimal one ($p>0.17$). 
These results demonstrates the potential of using thin films for studying the phase diagram of high-$T_c$ materials. A big advantage of thin films is that strain can be applied by the substrate in a relatively easy way, which open up for future studies of the relation between stain and superconductivity through measurements of $H_{c,2}$ by varying the thickness of the films going to the few nm scale.  

\hspace{0.2pt}

\section*{Acknowledgements}
This work was performed in part at Myfab Chalmers, and is supported by the Swedish Research Council (VR) under the projects 2018-04658 (F.L.), 2020-04945 (R.A.) and 2020-05184 (T.B.).

\bibliography{bibl}

\providecommand{\noopsort}[1]{}\providecommand{\singleletter}[1]{#1}%
\begin{thebibliography}{32}%
\makeatletter
\providecommand \@ifxundefined [1]{%
 \@ifx{#1\undefined}
}%
\providecommand \@ifnum [1]{%
 \ifnum #1\expandafter \@firstoftwo
 \else \expandafter \@secondoftwo
 \fi
}%
\providecommand \@ifx [1]{%
 \ifx #1\expandafter \@firstoftwo
 \else \expandafter \@secondoftwo
 \fi
}%
\providecommand \natexlab [1]{#1}%
\providecommand \enquote  [1]{``#1''}%
\providecommand \bibnamefont  [1]{#1}%
\providecommand \bibfnamefont [1]{#1}%
\providecommand \citenamefont [1]{#1}%
\providecommand \href@noop [0]{\@secondoftwo}%
\providecommand \href [0]{\begingroup \@sanitize@url \@href}%
\providecommand \@href[1]{\@@startlink{#1}\@@href}%
\providecommand \@@href[1]{\endgroup#1\@@endlink}%
\providecommand \@sanitize@url [0]{\catcode `\\12\catcode `\$12\catcode
  `\&12\catcode `\#12\catcode `\^12\catcode `\_12\catcode `\%12\relax}%
\providecommand \@@startlink[1]{}%
\providecommand \@@endlink[0]{}%
\providecommand \url  [0]{\begingroup\@sanitize@url \@url }%
\providecommand \@url [1]{\endgroup\@href {#1}{\urlprefix }}%
\providecommand \urlprefix  [0]{URL }%
\providecommand \Eprint [0]{\href }%
\providecommand \doibase [0]{https://doi.org/}%
\providecommand \selectlanguage [0]{\@gobble}%
\providecommand \bibinfo  [0]{\@secondoftwo}%
\providecommand \bibfield  [0]{\@secondoftwo}%
\providecommand \translation [1]{[#1]}%
\providecommand \BibitemOpen [0]{}%
\providecommand \bibitemStop [0]{}%
\providecommand \bibitemNoStop [0]{.\EOS\space}%
\providecommand \EOS [0]{\spacefactor3000\relax}%
\providecommand \BibitemShut  [1]{\csname bibitem#1\endcsname}%
\let\auto@bib@innerbib\@empty
\bibitem [{\citenamefont {Ghiringhelli}\ \emph {et~al.}(2012)\citenamefont
  {Ghiringhelli}, \citenamefont {Tacon}, \citenamefont {Minola}, \citenamefont
  {Blanco-Canosa}, \citenamefont {Mazzoli}, \citenamefont {Brookes},
  \citenamefont {Luca}, \citenamefont {Frano}, \citenamefont {Hawthorn},
  \citenamefont {He}, \citenamefont {Loew}, \citenamefont {Sala}, \citenamefont
  {Peets}, \citenamefont {Salluzzo}, \citenamefont {Schierle}, \citenamefont
  {Sutarto}, \citenamefont {Sawatzky}, \citenamefont {Weschke}, \citenamefont
  {Keimer},\ and\ \citenamefont {Braicovich}}]{Ghiringhelli2012}%
  \BibitemOpen
  \bibfield  {author} {\bibinfo {author} {\bibfnamefont {G.}~\bibnamefont
  {Ghiringhelli}}, \bibinfo {author} {\bibfnamefont {M.~L.}\ \bibnamefont
  {Tacon}}, \bibinfo {author} {\bibfnamefont {M.}~\bibnamefont {Minola}},
  \bibinfo {author} {\bibfnamefont {S.}~\bibnamefont {Blanco-Canosa}}, \bibinfo
  {author} {\bibfnamefont {C.}~\bibnamefont {Mazzoli}}, \bibinfo {author}
  {\bibfnamefont {N.}~\bibnamefont {Brookes}}, \bibinfo {author} {\bibfnamefont
  {G.~D.}\ \bibnamefont {Luca}}, \bibinfo {author} {\bibfnamefont
  {A.}~\bibnamefont {Frano}}, \bibinfo {author} {\bibfnamefont
  {D.}~\bibnamefont {Hawthorn}}, \bibinfo {author} {\bibfnamefont
  {F.}~\bibnamefont {He}}, \bibinfo {author} {\bibfnamefont {T.}~\bibnamefont
  {Loew}}, \bibinfo {author} {\bibfnamefont {M.~M.}\ \bibnamefont {Sala}},
  \bibinfo {author} {\bibfnamefont {D.}~\bibnamefont {Peets}}, \bibinfo
  {author} {\bibfnamefont {M.}~\bibnamefont {Salluzzo}}, \bibinfo {author}
  {\bibfnamefont {E.}~\bibnamefont {Schierle}}, \bibinfo {author}
  {\bibfnamefont {R.}~\bibnamefont {Sutarto}}, \bibinfo {author} {\bibfnamefont
  {G.}~\bibnamefont {Sawatzky}}, \bibinfo {author} {\bibfnamefont
  {E.}~\bibnamefont {Weschke}}, \bibinfo {author} {\bibfnamefont
  {B.}~\bibnamefont {Keimer}},\ and\ \bibinfo {author} {\bibfnamefont
  {L.}~\bibnamefont {Braicovich}},\ }\href@noop {} {\bibfield  {journal}
  {\bibinfo  {journal} {Science}\ }\textbf {\bibinfo {volume} {337}},\ \bibinfo
  {pages} {821} (\bibinfo {year} {2012})}\BibitemShut {NoStop}%
\bibitem [{\citenamefont {Chang}\ \emph
  {et~al.}(2012{\natexlab{a}})\citenamefont {Chang}, \citenamefont {Blackburn},
  \citenamefont {Holmes}, \citenamefont {Christensen}, \citenamefont {Larsen},
  \citenamefont {Mesot}, \citenamefont {Liang}, \citenamefont {Bonn},
  \citenamefont {Hardy}, \citenamefont {Watenphul}, \citenamefont
  {v.~Zimmermann}, \citenamefont {Forgan},\ and\ \citenamefont
  {Hayden}}]{Chang2012}%
  \BibitemOpen
  \bibfield  {author} {\bibinfo {author} {\bibfnamefont {J.}~\bibnamefont
  {Chang}}, \bibinfo {author} {\bibfnamefont {E.}~\bibnamefont {Blackburn}},
  \bibinfo {author} {\bibfnamefont {A.~T.}\ \bibnamefont {Holmes}}, \bibinfo
  {author} {\bibfnamefont {N.~B.}\ \bibnamefont {Christensen}}, \bibinfo
  {author} {\bibfnamefont {J.}~\bibnamefont {Larsen}}, \bibinfo {author}
  {\bibfnamefont {J.}~\bibnamefont {Mesot}}, \bibinfo {author} {\bibfnamefont
  {R.}~\bibnamefont {Liang}}, \bibinfo {author} {\bibfnamefont {D.~A.}\
  \bibnamefont {Bonn}}, \bibinfo {author} {\bibfnamefont {W.~N.}\ \bibnamefont
  {Hardy}}, \bibinfo {author} {\bibfnamefont {A.}~\bibnamefont {Watenphul}},
  \bibinfo {author} {\bibfnamefont {M.}~\bibnamefont {v.~Zimmermann}}, \bibinfo
  {author} {\bibfnamefont {E.~M.}\ \bibnamefont {Forgan}},\ and\ \bibinfo
  {author} {\bibfnamefont {S.~M.}\ \bibnamefont {Hayden}},\ }\href@noop {}
  {\bibfield  {journal} {\bibinfo  {journal} {Nat. Phys.}\ }\textbf {\bibinfo
  {volume} {8}},\ \bibinfo {pages} {871} (\bibinfo {year}
  {2012}{\natexlab{a}})}\BibitemShut {NoStop}%
\bibitem [{\citenamefont {Blanco-Canosa}\ \emph {et~al.}(2014)\citenamefont
  {Blanco-Canosa}, \citenamefont {Frano}, \citenamefont {Schierle},
  \citenamefont {Porras}, \citenamefont {Loew}, \citenamefont {Minola},
  \citenamefont {Bluschke}, \citenamefont {Weschke}, \citenamefont {Keimer},\
  and\ \citenamefont {Tacon}}]{Blanco2014}%
  \BibitemOpen
  \bibfield  {author} {\bibinfo {author} {\bibfnamefont {S.}~\bibnamefont
  {Blanco-Canosa}}, \bibinfo {author} {\bibfnamefont {A.}~\bibnamefont
  {Frano}}, \bibinfo {author} {\bibfnamefont {E.}~\bibnamefont {Schierle}},
  \bibinfo {author} {\bibfnamefont {J.}~\bibnamefont {Porras}}, \bibinfo
  {author} {\bibfnamefont {T.}~\bibnamefont {Loew}}, \bibinfo {author}
  {\bibfnamefont {M.}~\bibnamefont {Minola}}, \bibinfo {author} {\bibfnamefont
  {M.}~\bibnamefont {Bluschke}}, \bibinfo {author} {\bibfnamefont
  {E.}~\bibnamefont {Weschke}}, \bibinfo {author} {\bibfnamefont
  {B.}~\bibnamefont {Keimer}},\ and\ \bibinfo {author} {\bibfnamefont {M.~L.}\
  \bibnamefont {Tacon}},\ }\href@noop {} {\bibfield  {journal} {\bibinfo
  {journal} {Phys. Rev. B}\ }\textbf {\bibinfo {volume} {90}},\ \bibinfo
  {pages} {054513} (\bibinfo {year} {2014})}\BibitemShut {NoStop}%
\bibitem [{\citenamefont {Cyr-Choinière}\ \emph {et~al.}(2018)\citenamefont
  {Cyr-Choinière}, \citenamefont {LeBoeuf}, \citenamefont {Badoux},
  \citenamefont {Dufour-Beauséjour}, \citenamefont {Bonn}, \citenamefont
  {Hardy}, \citenamefont {Graf}, \citenamefont {Doiron-Leyraud},\ and\
  \citenamefont {Taillefer}}]{Cyrchoinere2018}%
  \BibitemOpen
  \bibfield  {author} {\bibinfo {author} {\bibfnamefont {O.}~\bibnamefont
  {Cyr-Choinière}}, \bibinfo {author} {\bibfnamefont {D.}~\bibnamefont
  {LeBoeuf}}, \bibinfo {author} {\bibfnamefont {S.}~\bibnamefont {Badoux}},
  \bibinfo {author} {\bibfnamefont {S.}~\bibnamefont {Dufour-Beauséjour}},
  \bibinfo {author} {\bibfnamefont {D.}~\bibnamefont {Bonn}}, \bibinfo {author}
  {\bibfnamefont {W.}~\bibnamefont {Hardy}}, \bibinfo {author} {\bibfnamefont
  {D.}~\bibnamefont {Graf}}, \bibinfo {author} {\bibfnamefont {N.}~\bibnamefont
  {Doiron-Leyraud}},\ and\ \bibinfo {author} {\bibfnamefont {L.}~\bibnamefont
  {Taillefer}},\ }\href@noop {} {\bibfield  {journal} {\bibinfo  {journal}
  {Phys. Rev. B}\ }\textbf {\bibinfo {volume} {98}},\ \bibinfo {pages} {064513}
  (\bibinfo {year} {2018})}\BibitemShut {NoStop}%
\bibitem [{\citenamefont {Arpaia}\ and\ \citenamefont
  {Ghiringhelli}(2021)}]{arpaia2021charge}%
  \BibitemOpen
  \bibfield  {author} {\bibinfo {author} {\bibfnamefont {R.}~\bibnamefont
  {Arpaia}}\ and\ \bibinfo {author} {\bibfnamefont {G.}~\bibnamefont
  {Ghiringhelli}},\ }\href@noop {} {\bibfield  {journal} {\bibinfo  {journal}
  {J. Phys. Soc. Jpn.}\ }\textbf {\bibinfo {volume} {90}},\ \bibinfo {pages}
  {111005} (\bibinfo {year} {2021})}\BibitemShut {NoStop}%
\bibitem [{\citenamefont {Ando}\ \emph {et~al.}(2004)\citenamefont {Ando},
  \citenamefont {Komiya}, \citenamefont {Segawa}, \citenamefont {Ono},\ and\
  \citenamefont {Kurita}}]{Ando2004}%
  \BibitemOpen
  \bibfield  {author} {\bibinfo {author} {\bibfnamefont {Y.}~\bibnamefont
  {Ando}}, \bibinfo {author} {\bibfnamefont {S.}~\bibnamefont {Komiya}},
  \bibinfo {author} {\bibfnamefont {K.}~\bibnamefont {Segawa}}, \bibinfo
  {author} {\bibfnamefont {S.}~\bibnamefont {Ono}},\ and\ \bibinfo {author}
  {\bibfnamefont {Y.}~\bibnamefont {Kurita}},\ }\href@noop {} {\bibfield
  {journal} {\bibinfo  {journal} {Phys. Rev. Lett.}\ }\textbf {\bibinfo
  {volume} {93}},\ \bibinfo {pages} {267001} (\bibinfo {year}
  {2004})}\BibitemShut {NoStop}%
\bibitem [{\citenamefont {Liang}\ \emph {et~al.}(2006)\citenamefont {Liang},
  \citenamefont {Bonn},\ and\ \citenamefont {Hardy}}]{Liang2006}%
  \BibitemOpen
  \bibfield  {author} {\bibinfo {author} {\bibfnamefont {R.}~\bibnamefont
  {Liang}}, \bibinfo {author} {\bibfnamefont {D.}~\bibnamefont {Bonn}},\ and\
  \bibinfo {author} {\bibfnamefont {W.}~\bibnamefont {Hardy}},\ }\href@noop {}
  {\bibfield  {journal} {\bibinfo  {journal} {Phys. Rev. B.}\ }\textbf
  {\bibinfo {volume} {73}},\ \bibinfo {pages} {180505(R)} (\bibinfo {year}
  {2006})}\BibitemShut {NoStop}%
\bibitem [{\citenamefont {Huecker}\ \emph {et~al.}(2014)\citenamefont
  {Huecker}, \citenamefont {Christensen}, \citenamefont {Holmes}, \citenamefont
  {Blackburn}, \citenamefont {Forgan}, \citenamefont {Liang}, \citenamefont
  {Bonn}, \citenamefont {Hardy}, \citenamefont {Gutowski}, \citenamefont
  {Zimmermann} \emph {et~al.}}]{huecker2014competing}%
  \BibitemOpen
  \bibfield  {author} {\bibinfo {author} {\bibfnamefont {M.}~\bibnamefont
  {Huecker}}, \bibinfo {author} {\bibfnamefont {N.~B.}\ \bibnamefont
  {Christensen}}, \bibinfo {author} {\bibfnamefont {A.}~\bibnamefont {Holmes}},
  \bibinfo {author} {\bibfnamefont {E.}~\bibnamefont {Blackburn}}, \bibinfo
  {author} {\bibfnamefont {E.~M.}\ \bibnamefont {Forgan}}, \bibinfo {author}
  {\bibfnamefont {R.}~\bibnamefont {Liang}}, \bibinfo {author} {\bibfnamefont
  {D.}~\bibnamefont {Bonn}}, \bibinfo {author} {\bibfnamefont {W.}~\bibnamefont
  {Hardy}}, \bibinfo {author} {\bibfnamefont {O.}~\bibnamefont {Gutowski}},
  \bibinfo {author} {\bibfnamefont {M.~v.}\ \bibnamefont {Zimmermann}}, \emph
  {et~al.},\ }\href@noop {} {\bibfield  {journal} {\bibinfo  {journal} {Phys.
  Rev. B}\ }\textbf {\bibinfo {volume} {90}},\ \bibinfo {pages} {054514}
  (\bibinfo {year} {2014})}\BibitemShut {NoStop}%
\bibitem [{\citenamefont {Ando}\ and\ \citenamefont
  {Segawa}(2002)}]{ando2002magnetoresistance}%
  \BibitemOpen
  \bibfield  {author} {\bibinfo {author} {\bibfnamefont {Y.}~\bibnamefont
  {Ando}}\ and\ \bibinfo {author} {\bibfnamefont {K.}~\bibnamefont {Segawa}},\
  }\href@noop {} {\bibfield  {journal} {\bibinfo  {journal} {Phys. Rev. Lett.}\
  }\textbf {\bibinfo {volume} {88}},\ \bibinfo {pages} {167005} (\bibinfo
  {year} {2002})}\BibitemShut {NoStop}%
\bibitem [{\citenamefont {Ramshaw}\ \emph {et~al.}(2012)\citenamefont
  {Ramshaw}, \citenamefont {Day}, \citenamefont {Vignolle}, \citenamefont
  {LeBoeuf}, \citenamefont {Dosanjh}, \citenamefont {Proust}, \citenamefont
  {Taillefer}, \citenamefont {Liang}, \citenamefont {Hardy},\ and\
  \citenamefont {Bonn}}]{ramshaw2012vortex}%
  \BibitemOpen
  \bibfield  {author} {\bibinfo {author} {\bibfnamefont {B.}~\bibnamefont
  {Ramshaw}}, \bibinfo {author} {\bibfnamefont {J.}~\bibnamefont {Day}},
  \bibinfo {author} {\bibfnamefont {B.}~\bibnamefont {Vignolle}}, \bibinfo
  {author} {\bibfnamefont {D.}~\bibnamefont {LeBoeuf}}, \bibinfo {author}
  {\bibfnamefont {P.}~\bibnamefont {Dosanjh}}, \bibinfo {author} {\bibfnamefont
  {C.}~\bibnamefont {Proust}}, \bibinfo {author} {\bibfnamefont
  {L.}~\bibnamefont {Taillefer}}, \bibinfo {author} {\bibfnamefont
  {R.}~\bibnamefont {Liang}}, \bibinfo {author} {\bibfnamefont
  {W.}~\bibnamefont {Hardy}},\ and\ \bibinfo {author} {\bibfnamefont
  {D.}~\bibnamefont {Bonn}},\ }\href@noop {} {\bibfield  {journal} {\bibinfo
  {journal} {Phys. Rev. B}\ }\textbf {\bibinfo {volume} {86}},\ \bibinfo
  {pages} {174501} (\bibinfo {year} {2012})}\BibitemShut {NoStop}%
\bibitem [{\citenamefont {Chang}\ \emph
  {et~al.}(2012{\natexlab{b}})\citenamefont {Chang}, \citenamefont
  {Doiron-Leyraud}, \citenamefont {Cyr-Choiniere}, \citenamefont
  {Grissonnanche}, \citenamefont {Lalibert{\'e}}, \citenamefont {Hassinger},
  \citenamefont {Reid}, \citenamefont {Daou}, \citenamefont {Pyon},
  \citenamefont {Takayama} \emph {et~al.}}]{chang2012decrease}%
  \BibitemOpen
  \bibfield  {author} {\bibinfo {author} {\bibfnamefont {J.}~\bibnamefont
  {Chang}}, \bibinfo {author} {\bibfnamefont {N.}~\bibnamefont
  {Doiron-Leyraud}}, \bibinfo {author} {\bibfnamefont {O.}~\bibnamefont
  {Cyr-Choiniere}}, \bibinfo {author} {\bibfnamefont {G.}~\bibnamefont
  {Grissonnanche}}, \bibinfo {author} {\bibfnamefont {F.}~\bibnamefont
  {Lalibert{\'e}}}, \bibinfo {author} {\bibfnamefont {E.}~\bibnamefont
  {Hassinger}}, \bibinfo {author} {\bibfnamefont {J.-P.}\ \bibnamefont {Reid}},
  \bibinfo {author} {\bibfnamefont {R.}~\bibnamefont {Daou}}, \bibinfo {author}
  {\bibfnamefont {S.}~\bibnamefont {Pyon}}, \bibinfo {author} {\bibfnamefont
  {T.}~\bibnamefont {Takayama}}, \emph {et~al.},\ }\href@noop {} {\bibfield
  {journal} {\bibinfo  {journal} {Nat. Phys.}\ }\textbf {\bibinfo {volume}
  {8}},\ \bibinfo {pages} {751} (\bibinfo {year}
  {2012}{\natexlab{b}})}\BibitemShut {NoStop}%
\bibitem [{\citenamefont {Grissonnanche}\ \emph {et~al.}(2014)\citenamefont
  {Grissonnanche}, \citenamefont {Cyr-Choini{\`e}re}, \citenamefont
  {Lalibert{\'e}}, \citenamefont {Ren{\'e}~de Cotret}, \citenamefont
  {Juneau-Fecteau}, \citenamefont {Dufour-Beaus{\'e}jour}, \citenamefont
  {Delage}, \citenamefont {LeBoeuf}, \citenamefont {Chang}, \citenamefont
  {Ramshaw} \emph {et~al.}}]{grissonnanche2014direct}%
  \BibitemOpen
  \bibfield  {author} {\bibinfo {author} {\bibfnamefont {G.}~\bibnamefont
  {Grissonnanche}}, \bibinfo {author} {\bibfnamefont {O.}~\bibnamefont
  {Cyr-Choini{\`e}re}}, \bibinfo {author} {\bibfnamefont {F.}~\bibnamefont
  {Lalibert{\'e}}}, \bibinfo {author} {\bibfnamefont {S.}~\bibnamefont
  {Ren{\'e}~de Cotret}}, \bibinfo {author} {\bibfnamefont {A.}~\bibnamefont
  {Juneau-Fecteau}}, \bibinfo {author} {\bibfnamefont {S.}~\bibnamefont
  {Dufour-Beaus{\'e}jour}}, \bibinfo {author} {\bibfnamefont {M.-E.}\
  \bibnamefont {Delage}}, \bibinfo {author} {\bibfnamefont {D.}~\bibnamefont
  {LeBoeuf}}, \bibinfo {author} {\bibfnamefont {J.}~\bibnamefont {Chang}},
  \bibinfo {author} {\bibfnamefont {B.}~\bibnamefont {Ramshaw}}, \emph
  {et~al.},\ }\href@noop {} {\bibfield  {journal} {\bibinfo  {journal} {Nat.
  Commun.}\ }\textbf {\bibinfo {volume} {5}},\ \bibinfo {pages} {1} (\bibinfo
  {year} {2014})}\BibitemShut {NoStop}%
\bibitem [{\citenamefont {Ando}\ \emph {et~al.}(1999)\citenamefont {Ando},
  \citenamefont {Boebinger}, \citenamefont {Passner}, \citenamefont
  {Schneemeyer}, \citenamefont {Kimura}, \citenamefont {Okuya}, \citenamefont
  {Watauchi}, \citenamefont {Shimoyama}, \citenamefont {Kishio}, \citenamefont
  {Tamasaku} \emph {et~al.}}]{ando1999resistive}%
  \BibitemOpen
  \bibfield  {author} {\bibinfo {author} {\bibfnamefont {Y.}~\bibnamefont
  {Ando}}, \bibinfo {author} {\bibfnamefont {G.~S.}\ \bibnamefont {Boebinger}},
  \bibinfo {author} {\bibfnamefont {A.}~\bibnamefont {Passner}}, \bibinfo
  {author} {\bibfnamefont {L.}~\bibnamefont {Schneemeyer}}, \bibinfo {author}
  {\bibfnamefont {T.}~\bibnamefont {Kimura}}, \bibinfo {author} {\bibfnamefont
  {M.}~\bibnamefont {Okuya}}, \bibinfo {author} {\bibfnamefont
  {S.}~\bibnamefont {Watauchi}}, \bibinfo {author} {\bibfnamefont
  {J.}~\bibnamefont {Shimoyama}}, \bibinfo {author} {\bibfnamefont
  {K.}~\bibnamefont {Kishio}}, \bibinfo {author} {\bibfnamefont
  {K.}~\bibnamefont {Tamasaku}}, \emph {et~al.},\ }\href@noop {} {\bibfield
  {journal} {\bibinfo  {journal} {Phys. Rev. B}\ }\textbf {\bibinfo {volume}
  {60}},\ \bibinfo {pages} {12475} (\bibinfo {year} {1999})}\BibitemShut
  {NoStop}%
\bibitem [{\citenamefont {Wang}\ \emph {et~al.}(2003)\citenamefont {Wang},
  \citenamefont {Ono}, \citenamefont {Onose}, \citenamefont {Gu}, \citenamefont
  {Ando}, \citenamefont {Tokura}, \citenamefont {Uchida},\ and\ \citenamefont
  {Ong}}]{wang2003dependence}%
  \BibitemOpen
  \bibfield  {author} {\bibinfo {author} {\bibfnamefont {Y.}~\bibnamefont
  {Wang}}, \bibinfo {author} {\bibfnamefont {S.}~\bibnamefont {Ono}}, \bibinfo
  {author} {\bibfnamefont {Y.}~\bibnamefont {Onose}}, \bibinfo {author}
  {\bibfnamefont {G.}~\bibnamefont {Gu}}, \bibinfo {author} {\bibfnamefont
  {Y.}~\bibnamefont {Ando}}, \bibinfo {author} {\bibfnamefont {Y.}~\bibnamefont
  {Tokura}}, \bibinfo {author} {\bibfnamefont {S.}~\bibnamefont {Uchida}},\
  and\ \bibinfo {author} {\bibfnamefont {N.}~\bibnamefont {Ong}},\ }\href@noop
  {} {\bibfield  {journal} {\bibinfo  {journal} {Science}\ }\textbf {\bibinfo
  {volume} {299}},\ \bibinfo {pages} {86} (\bibinfo {year} {2003})}\BibitemShut
  {NoStop}%
\bibitem [{\citenamefont {Kim}\ \emph {et~al.}(2018)\citenamefont {Kim},
  \citenamefont {Souliou}, \citenamefont {Barber}, \citenamefont
  {Lefran{\c{c}}ois}, \citenamefont {Minola}, \citenamefont {Tortora},
  \citenamefont {Heid}, \citenamefont {Nandi}, \citenamefont {Borzi},
  \citenamefont {Garbarino} \emph {et~al.}}]{kim2018uniaxial}%
  \BibitemOpen
  \bibfield  {author} {\bibinfo {author} {\bibfnamefont {H.-H.}\ \bibnamefont
  {Kim}}, \bibinfo {author} {\bibfnamefont {S.}~\bibnamefont {Souliou}},
  \bibinfo {author} {\bibfnamefont {M.}~\bibnamefont {Barber}}, \bibinfo
  {author} {\bibfnamefont {E.}~\bibnamefont {Lefran{\c{c}}ois}}, \bibinfo
  {author} {\bibfnamefont {M.}~\bibnamefont {Minola}}, \bibinfo {author}
  {\bibfnamefont {M.}~\bibnamefont {Tortora}}, \bibinfo {author} {\bibfnamefont
  {R.}~\bibnamefont {Heid}}, \bibinfo {author} {\bibfnamefont {N.}~\bibnamefont
  {Nandi}}, \bibinfo {author} {\bibfnamefont {R.~A.}\ \bibnamefont {Borzi}},
  \bibinfo {author} {\bibfnamefont {G.}~\bibnamefont {Garbarino}}, \emph
  {et~al.},\ }\href@noop {} {\bibfield  {journal} {\bibinfo  {journal}
  {Science}\ }\textbf {\bibinfo {volume} {362}},\ \bibinfo {pages} {1040}
  (\bibinfo {year} {2018})}\BibitemShut {NoStop}%
\bibitem [{\citenamefont {Kim}\ \emph {et~al.}(2021)\citenamefont {Kim},
  \citenamefont {Lefran{\c{c}}ois}, \citenamefont {Kummer}, \citenamefont
  {Fumagalli}, \citenamefont {Brookes}, \citenamefont {Betto}, \citenamefont
  {Nakata}, \citenamefont {Tortora}, \citenamefont {Porras}, \citenamefont
  {Loew} \emph {et~al.}}]{kim2021charge}%
  \BibitemOpen
  \bibfield  {author} {\bibinfo {author} {\bibfnamefont {H.-H.}\ \bibnamefont
  {Kim}}, \bibinfo {author} {\bibfnamefont {E.}~\bibnamefont
  {Lefran{\c{c}}ois}}, \bibinfo {author} {\bibfnamefont {K.}~\bibnamefont
  {Kummer}}, \bibinfo {author} {\bibfnamefont {R.}~\bibnamefont {Fumagalli}},
  \bibinfo {author} {\bibfnamefont {N.}~\bibnamefont {Brookes}}, \bibinfo
  {author} {\bibfnamefont {D.}~\bibnamefont {Betto}}, \bibinfo {author}
  {\bibfnamefont {S.}~\bibnamefont {Nakata}}, \bibinfo {author} {\bibfnamefont
  {M.}~\bibnamefont {Tortora}}, \bibinfo {author} {\bibfnamefont
  {J.}~\bibnamefont {Porras}}, \bibinfo {author} {\bibfnamefont
  {T.}~\bibnamefont {Loew}}, \emph {et~al.},\ }\href@noop {} {\bibfield
  {journal} {\bibinfo  {journal} {Phys. Rev. Lett.}\ }\textbf {\bibinfo
  {volume} {126}},\ \bibinfo {pages} {037002} (\bibinfo {year}
  {2021})}\BibitemShut {NoStop}%
\bibitem [{\citenamefont {Souliou}\ \emph {et~al.}(2018)\citenamefont
  {Souliou}, \citenamefont {Gretarsson}, \citenamefont {Garbarino},
  \citenamefont {Bosak}, \citenamefont {Porras}, \citenamefont {Loew},
  \citenamefont {Keimer},\ and\ \citenamefont {Tacon}}]{Souliou2018}%
  \BibitemOpen
  \bibfield  {author} {\bibinfo {author} {\bibfnamefont {S.}~\bibnamefont
  {Souliou}}, \bibinfo {author} {\bibfnamefont {H.}~\bibnamefont {Gretarsson}},
  \bibinfo {author} {\bibfnamefont {G.}~\bibnamefont {Garbarino}}, \bibinfo
  {author} {\bibfnamefont {A.}~\bibnamefont {Bosak}}, \bibinfo {author}
  {\bibfnamefont {J.}~\bibnamefont {Porras}}, \bibinfo {author} {\bibfnamefont
  {T.}~\bibnamefont {Loew}}, \bibinfo {author} {\bibfnamefont {B.}~\bibnamefont
  {Keimer}},\ and\ \bibinfo {author} {\bibfnamefont {M.~L.}\ \bibnamefont
  {Tacon}},\ }\href@noop {} {\bibfield  {journal} {\bibinfo  {journal} {Phys.
  Rev. B}\ }\textbf {\bibinfo {volume} {97}},\ \bibinfo {pages} {020503(R)}
  (\bibinfo {year} {2018})}\BibitemShut {NoStop}%
\bibitem [{\citenamefont {Choi}\ \emph {et~al.}(2022)\citenamefont {Choi},
  \citenamefont {Wang}, \citenamefont {J{\"o}hr}, \citenamefont {Christensen},
  \citenamefont {K{\"u}spert}, \citenamefont {Bucher}, \citenamefont
  {Biscette}, \citenamefont {Fischer}, \citenamefont {H{\"u}cker},
  \citenamefont {Kurosawa} \emph {et~al.}}]{choi2022unveiling}%
  \BibitemOpen
  \bibfield  {author} {\bibinfo {author} {\bibfnamefont {J.}~\bibnamefont
  {Choi}}, \bibinfo {author} {\bibfnamefont {Q.}~\bibnamefont {Wang}}, \bibinfo
  {author} {\bibfnamefont {S.}~\bibnamefont {J{\"o}hr}}, \bibinfo {author}
  {\bibfnamefont {N.}~\bibnamefont {Christensen}}, \bibinfo {author}
  {\bibfnamefont {J.}~\bibnamefont {K{\"u}spert}}, \bibinfo {author}
  {\bibfnamefont {D.}~\bibnamefont {Bucher}}, \bibinfo {author} {\bibfnamefont
  {D.}~\bibnamefont {Biscette}}, \bibinfo {author} {\bibfnamefont
  {M.}~\bibnamefont {Fischer}}, \bibinfo {author} {\bibfnamefont
  {M.}~\bibnamefont {H{\"u}cker}}, \bibinfo {author} {\bibfnamefont
  {T.}~\bibnamefont {Kurosawa}}, \emph {et~al.},\ }\bibfield  {title} {\bibinfo
  {title} {Unveiling unequivocal charge stripe order in a prototypical cuprate
  superconductor},\ }\href@noop {} {\bibfield  {journal} {\bibinfo  {journal}
  {Phys. Rev. Lett.}\ }\textbf {\bibinfo {volume} {128}},\ \bibinfo {pages}
  {207002} (\bibinfo {year} {2022})}\BibitemShut {NoStop}%
\bibitem [{\citenamefont {Arpaia}\ \emph
  {et~al.}(2019{\natexlab{a}})\citenamefont {Arpaia}, \citenamefont
  {Andersson}, \citenamefont {Kalaboukhov}, \citenamefont {Schröder},
  \citenamefont {Trabaldo}, \citenamefont {Ciancio}, \citenamefont {Dražic},
  \citenamefont {Orgiani}, \citenamefont {Bauch},\ and\ \citenamefont
  {Lombardi}}]{arpaia2019}%
  \BibitemOpen
  \bibfield  {author} {\bibinfo {author} {\bibfnamefont {R.}~\bibnamefont
  {Arpaia}}, \bibinfo {author} {\bibfnamefont {E.}~\bibnamefont {Andersson}},
  \bibinfo {author} {\bibfnamefont {A.}~\bibnamefont {Kalaboukhov}}, \bibinfo
  {author} {\bibfnamefont {E.}~\bibnamefont {Schröder}}, \bibinfo {author}
  {\bibfnamefont {E.}~\bibnamefont {Trabaldo}}, \bibinfo {author}
  {\bibfnamefont {R.}~\bibnamefont {Ciancio}}, \bibinfo {author} {\bibfnamefont
  {G.}~\bibnamefont {Dražic}}, \bibinfo {author} {\bibfnamefont
  {P.}~\bibnamefont {Orgiani}}, \bibinfo {author} {\bibfnamefont
  {T.}~\bibnamefont {Bauch}},\ and\ \bibinfo {author} {\bibfnamefont
  {F.}~\bibnamefont {Lombardi}},\ }\href@noop {} {\bibfield  {journal}
  {\bibinfo  {journal} {Phys. Rev. Materials}\ }\textbf {\bibinfo {volume}
  {3}},\ \bibinfo {pages} {114804} (\bibinfo {year}
  {2019}{\natexlab{a}})}\BibitemShut {NoStop}%
\bibitem [{\citenamefont {Wahlberg}\ \emph {et~al.}(2021)\citenamefont
  {Wahlberg}, \citenamefont {Arpaia}, \citenamefont {Seibold}, \citenamefont
  {Rossi}, \citenamefont {Fumagalli}, \citenamefont {Trabaldo}, \citenamefont
  {Brookes}, \citenamefont {Braicovich}, \citenamefont {Caprara}, \citenamefont
  {Gran} \emph {et~al.}}]{wahlberg2021}%
  \BibitemOpen
  \bibfield  {author} {\bibinfo {author} {\bibfnamefont {E.}~\bibnamefont
  {Wahlberg}}, \bibinfo {author} {\bibfnamefont {R.}~\bibnamefont {Arpaia}},
  \bibinfo {author} {\bibfnamefont {G.}~\bibnamefont {Seibold}}, \bibinfo
  {author} {\bibfnamefont {M.}~\bibnamefont {Rossi}}, \bibinfo {author}
  {\bibfnamefont {R.}~\bibnamefont {Fumagalli}}, \bibinfo {author}
  {\bibfnamefont {E.}~\bibnamefont {Trabaldo}}, \bibinfo {author}
  {\bibfnamefont {N.~B.}\ \bibnamefont {Brookes}}, \bibinfo {author}
  {\bibfnamefont {L.}~\bibnamefont {Braicovich}}, \bibinfo {author}
  {\bibfnamefont {S.}~\bibnamefont {Caprara}}, \bibinfo {author} {\bibfnamefont
  {U.}~\bibnamefont {Gran}}, \emph {et~al.},\ }\href@noop {} {\bibfield
  {journal} {\bibinfo  {journal} {Science}\ }\textbf {\bibinfo {volume}
  {373}},\ \bibinfo {pages} {1506} (\bibinfo {year} {2021})}\BibitemShut
  {NoStop}%
\bibitem [{\citenamefont {Arpaia}\ \emph {et~al.}(2018)\citenamefont {Arpaia},
  \citenamefont {Andersson}, \citenamefont {Trabaldo}, \citenamefont {Bauch},\
  and\ \citenamefont {Lombardi}}]{arpaia2018}%
  \BibitemOpen
  \bibfield  {author} {\bibinfo {author} {\bibfnamefont {R.}~\bibnamefont
  {Arpaia}}, \bibinfo {author} {\bibfnamefont {E.}~\bibnamefont {Andersson}},
  \bibinfo {author} {\bibfnamefont {E.}~\bibnamefont {Trabaldo}}, \bibinfo
  {author} {\bibfnamefont {T.}~\bibnamefont {Bauch}},\ and\ \bibinfo {author}
  {\bibfnamefont {F.}~\bibnamefont {Lombardi}},\ }\href@noop {} {\bibfield
  {journal} {\bibinfo  {journal} {Phys. Rev. Materials}\ }\textbf {\bibinfo
  {volume} {2}},\ \bibinfo {pages} {024820} (\bibinfo {year}
  {2018})}\BibitemShut {NoStop}%
\bibitem [{\citenamefont {Baghdadi}\ \emph {et~al.}(2014)\citenamefont
  {Baghdadi}, \citenamefont {Arpaia}, \citenamefont {Bauch},\ and\
  \citenamefont {Lombardi}}]{baghdadi2014toward}%
  \BibitemOpen
  \bibfield  {author} {\bibinfo {author} {\bibfnamefont {R.}~\bibnamefont
  {Baghdadi}}, \bibinfo {author} {\bibfnamefont {R.}~\bibnamefont {Arpaia}},
  \bibinfo {author} {\bibfnamefont {T.}~\bibnamefont {Bauch}},\ and\ \bibinfo
  {author} {\bibfnamefont {F.}~\bibnamefont {Lombardi}},\ }\href@noop {}
  {\bibfield  {journal} {\bibinfo  {journal} {IEEE Trans. Appl. Supercond.}\
  }\textbf {\bibinfo {volume} {25}},\ \bibinfo {pages} {1} (\bibinfo {year}
  {2014})}\BibitemShut {NoStop}%
\bibitem [{\citenamefont {Jorgensen}\ \emph {et~al.}(1990)\citenamefont
  {Jorgensen}, \citenamefont {Veal}, \citenamefont {Paulikas}, \citenamefont
  {Nowicki}, \citenamefont {Crabtree}, \citenamefont {Claus},\ and\
  \citenamefont {Kwok}}]{Jorgensen1990}%
  \BibitemOpen
  \bibfield  {author} {\bibinfo {author} {\bibfnamefont {J.}~\bibnamefont
  {Jorgensen}}, \bibinfo {author} {\bibfnamefont {B.}~\bibnamefont {Veal}},
  \bibinfo {author} {\bibfnamefont {A.}~\bibnamefont {Paulikas}}, \bibinfo
  {author} {\bibfnamefont {L.}~\bibnamefont {Nowicki}}, \bibinfo {author}
  {\bibfnamefont {G.}~\bibnamefont {Crabtree}}, \bibinfo {author}
  {\bibfnamefont {H.}~\bibnamefont {Claus}},\ and\ \bibinfo {author}
  {\bibfnamefont {W.}~\bibnamefont {Kwok}},\ }\href@noop {} {\bibfield
  {journal} {\bibinfo  {journal} {Phys. Rev. B.}\ }\textbf {\bibinfo {volume}
  {41}},\ \bibinfo {pages} {1863} (\bibinfo {year} {1990})}\BibitemShut
  {NoStop}%
\bibitem [{\citenamefont {Gagnon}\ \emph {et~al.}(1994)\citenamefont {Gagnon},
  \citenamefont {Lupien},\ and\ \citenamefont {Taillefer}}]{gagnon1994t}%
  \BibitemOpen
  \bibfield  {author} {\bibinfo {author} {\bibfnamefont {R.}~\bibnamefont
  {Gagnon}}, \bibinfo {author} {\bibfnamefont {C.}~\bibnamefont {Lupien}},\
  and\ \bibinfo {author} {\bibfnamefont {L.}~\bibnamefont {Taillefer}},\
  }\href@noop {} {\bibfield  {journal} {\bibinfo  {journal} {Phys. Rev. B}\
  }\textbf {\bibinfo {volume} {50}},\ \bibinfo {pages} {3458} (\bibinfo {year}
  {1994})}\BibitemShut {NoStop}%
\bibitem [{\citenamefont {Ando}\ \emph {et~al.}(2002)\citenamefont {Ando},
  \citenamefont {Segawa}, \citenamefont {Komiya},\ and\ \citenamefont
  {Lavrov}}]{Ando2002}%
  \BibitemOpen
  \bibfield  {author} {\bibinfo {author} {\bibfnamefont {Y.}~\bibnamefont
  {Ando}}, \bibinfo {author} {\bibfnamefont {K.}~\bibnamefont {Segawa}},
  \bibinfo {author} {\bibfnamefont {S.}~\bibnamefont {Komiya}},\ and\ \bibinfo
  {author} {\bibfnamefont {A.}~\bibnamefont {Lavrov}},\ }\href@noop {}
  {\bibfield  {journal} {\bibinfo  {journal} {Phys. Rev. Lett.}\ }\textbf
  {\bibinfo {volume} {88}},\ \bibinfo {pages} {137005} (\bibinfo {year}
  {2002})}\BibitemShut {NoStop}%
\bibitem [{\citenamefont {Andersson}\ \emph {et~al.}(2020)\citenamefont
  {Andersson}, \citenamefont {Arpaia}, \citenamefont {Trabaldo}, \citenamefont
  {Bauch},\ and\ \citenamefont {Lombardi}}]{Andersson2020}%
  \BibitemOpen
  \bibfield  {author} {\bibinfo {author} {\bibfnamefont {E.}~\bibnamefont
  {Andersson}}, \bibinfo {author} {\bibfnamefont {R.}~\bibnamefont {Arpaia}},
  \bibinfo {author} {\bibfnamefont {E.}~\bibnamefont {Trabaldo}}, \bibinfo
  {author} {\bibfnamefont {T.}~\bibnamefont {Bauch}},\ and\ \bibinfo {author}
  {\bibfnamefont {F.}~\bibnamefont {Lombardi}},\ }\href@noop {} {\bibfield
  {journal} {\bibinfo  {journal} {Supercond. Sci. Technol.}\ }\textbf {\bibinfo
  {volume} {33}},\ \bibinfo {pages} {064002} (\bibinfo {year}
  {2020})}\BibitemShut {NoStop}%
\bibitem [{\citenamefont {Keimer}\ \emph {et~al.}(2015)\citenamefont {Keimer},
  \citenamefont {Kivelson}, \citenamefont {Norman}, \citenamefont {Uchida},\
  and\ \citenamefont {Zaanen}}]{Keimer2015}%
  \BibitemOpen
  \bibfield  {author} {\bibinfo {author} {\bibfnamefont {B.}~\bibnamefont
  {Keimer}}, \bibinfo {author} {\bibfnamefont {S.}~\bibnamefont {Kivelson}},
  \bibinfo {author} {\bibfnamefont {M.}~\bibnamefont {Norman}}, \bibinfo
  {author} {\bibfnamefont {S.}~\bibnamefont {Uchida}},\ and\ \bibinfo {author}
  {\bibfnamefont {J.}~\bibnamefont {Zaanen}},\ }\href@noop {} {\bibfield
  {journal} {\bibinfo  {journal} {Nature}\ }\textbf {\bibinfo {volume} {518}},\
  \bibinfo {pages} {179} (\bibinfo {year} {2015})}\BibitemShut {NoStop}%
\bibitem [{\citenamefont {Castellani}\ \emph {et~al.}(1996)\citenamefont
  {Castellani}, \citenamefont {Di~Castro},\ and\ \citenamefont
  {Grilli}}]{castellani1996non}%
  \BibitemOpen
  \bibfield  {author} {\bibinfo {author} {\bibfnamefont {C.}~\bibnamefont
  {Castellani}}, \bibinfo {author} {\bibfnamefont {C.}~\bibnamefont
  {Di~Castro}},\ and\ \bibinfo {author} {\bibfnamefont {M.}~\bibnamefont
  {Grilli}},\ }\href@noop {} {\bibfield  {journal} {\bibinfo  {journal} {Z.
  Phys. B}\ }\textbf {\bibinfo {volume} {103}},\ \bibinfo {pages} {137}
  (\bibinfo {year} {1996})}\BibitemShut {NoStop}%
\bibitem [{\citenamefont {Arpaia}\ \emph
  {et~al.}(2019{\natexlab{b}})\citenamefont {Arpaia}, \citenamefont {Caprara},
  \citenamefont {Fumagalli}, \citenamefont {De~Vecchi}, \citenamefont {Peng},
  \citenamefont {Andersson}, \citenamefont {Betto}, \citenamefont {De~Luca},
  \citenamefont {Brookes}, \citenamefont {Lombardi}, \citenamefont {Salluzzo},
  \citenamefont {Braicovich}, \citenamefont {Di~Castro}, \citenamefont
  {Grilli},\ and\ \citenamefont {Ghiringhelli}}]{arpaia2019dynamical}%
  \BibitemOpen
  \bibfield  {author} {\bibinfo {author} {\bibfnamefont {R.}~\bibnamefont
  {Arpaia}}, \bibinfo {author} {\bibfnamefont {S.}~\bibnamefont {Caprara}},
  \bibinfo {author} {\bibfnamefont {R.}~\bibnamefont {Fumagalli}}, \bibinfo
  {author} {\bibfnamefont {G.}~\bibnamefont {De~Vecchi}}, \bibinfo {author}
  {\bibfnamefont {Y.~Y.}\ \bibnamefont {Peng}}, \bibinfo {author}
  {\bibfnamefont {E.}~\bibnamefont {Andersson}}, \bibinfo {author}
  {\bibfnamefont {D.}~\bibnamefont {Betto}}, \bibinfo {author} {\bibfnamefont
  {G.~M.}\ \bibnamefont {De~Luca}}, \bibinfo {author} {\bibfnamefont {N.~B.}\
  \bibnamefont {Brookes}}, \bibinfo {author} {\bibfnamefont {F.}~\bibnamefont
  {Lombardi}}, \bibinfo {author} {\bibfnamefont {M.}~\bibnamefont {Salluzzo}},
  \bibinfo {author} {\bibfnamefont {L.}~\bibnamefont {Braicovich}}, \bibinfo
  {author} {\bibfnamefont {C.}~\bibnamefont {Di~Castro}}, \bibinfo {author}
  {\bibfnamefont {M.}~\bibnamefont {Grilli}},\ and\ \bibinfo {author}
  {\bibfnamefont {G.}~\bibnamefont {Ghiringhelli}},\ }\href@noop {} {\bibfield
  {journal} {\bibinfo  {journal} {Science}\ }\textbf {\bibinfo {volume}
  {365}},\ \bibinfo {pages} {906} (\bibinfo {year}
  {2019}{\natexlab{b}})}\BibitemShut {NoStop}%
\bibitem [{\citenamefont {Ramshaw}\ \emph {et~al.}(2015)\citenamefont
  {Ramshaw}, \citenamefont {Sebastian}, \citenamefont {McDonald}, \citenamefont
  {Day}, \citenamefont {Tan}, \citenamefont {Zhu}, \citenamefont {Betts},
  \citenamefont {Liang}, \citenamefont {Bonn}, \citenamefont {Hardy} \emph
  {et~al.}}]{ramshaw2015quasiparticle}%
  \BibitemOpen
  \bibfield  {author} {\bibinfo {author} {\bibfnamefont {B.}~\bibnamefont
  {Ramshaw}}, \bibinfo {author} {\bibfnamefont {S.}~\bibnamefont {Sebastian}},
  \bibinfo {author} {\bibfnamefont {R.}~\bibnamefont {McDonald}}, \bibinfo
  {author} {\bibfnamefont {J.}~\bibnamefont {Day}}, \bibinfo {author}
  {\bibfnamefont {B.}~\bibnamefont {Tan}}, \bibinfo {author} {\bibfnamefont
  {Z.}~\bibnamefont {Zhu}}, \bibinfo {author} {\bibfnamefont {J.}~\bibnamefont
  {Betts}}, \bibinfo {author} {\bibfnamefont {R.}~\bibnamefont {Liang}},
  \bibinfo {author} {\bibfnamefont {D.}~\bibnamefont {Bonn}}, \bibinfo {author}
  {\bibfnamefont {W.}~\bibnamefont {Hardy}}, \emph {et~al.},\ }\href@noop {}
  {\bibfield  {journal} {\bibinfo  {journal} {Science}\ }\textbf {\bibinfo
  {volume} {348}},\ \bibinfo {pages} {317} (\bibinfo {year}
  {2015})}\BibitemShut {NoStop}%
\bibitem [{\citenamefont {Michon}\ \emph {et~al.}(2019)\citenamefont {Michon},
  \citenamefont {Girod}, \citenamefont {Badoux}, \citenamefont
  {Ka{\v{c}}mar{\v{c}}{\'\i}k}, \citenamefont {Ma}, \citenamefont {Dragomir},
  \citenamefont {Dabkowska}, \citenamefont {Gaulin}, \citenamefont {Zhou},
  \citenamefont {Pyon} \emph {et~al.}}]{michon2019thermodynamic}%
  \BibitemOpen
  \bibfield  {author} {\bibinfo {author} {\bibfnamefont {B.}~\bibnamefont
  {Michon}}, \bibinfo {author} {\bibfnamefont {C.}~\bibnamefont {Girod}},
  \bibinfo {author} {\bibfnamefont {S.}~\bibnamefont {Badoux}}, \bibinfo
  {author} {\bibfnamefont {J.}~\bibnamefont {Ka{\v{c}}mar{\v{c}}{\'\i}k}},
  \bibinfo {author} {\bibfnamefont {Q.}~\bibnamefont {Ma}}, \bibinfo {author}
  {\bibfnamefont {M.}~\bibnamefont {Dragomir}}, \bibinfo {author}
  {\bibfnamefont {H.}~\bibnamefont {Dabkowska}}, \bibinfo {author}
  {\bibfnamefont {B.}~\bibnamefont {Gaulin}}, \bibinfo {author} {\bibfnamefont
  {J.-S.}\ \bibnamefont {Zhou}}, \bibinfo {author} {\bibfnamefont
  {S.}~\bibnamefont {Pyon}}, \emph {et~al.},\ }\href@noop {} {\bibfield
  {journal} {\bibinfo  {journal} {Nature}\ }\textbf {\bibinfo {volume} {567}},\
  \bibinfo {pages} {218} (\bibinfo {year} {2019})}\BibitemShut {NoStop}%
\bibitem [{\citenamefont {LeBoeuf}\ \emph {et~al.}(2007)\citenamefont
  {LeBoeuf}, \citenamefont {Doiron-Leyraud}, \citenamefont {Levallois},
  \citenamefont {Daou}, \citenamefont {Bonnemaison}, \citenamefont {Hussey},
  \citenamefont {Balicas}, \citenamefont {Ramshaw}, \citenamefont {Liang},
  \citenamefont {Bonn} \emph {et~al.}}]{leboeuf2007electron}%
  \BibitemOpen
  \bibfield  {author} {\bibinfo {author} {\bibfnamefont {D.}~\bibnamefont
  {LeBoeuf}}, \bibinfo {author} {\bibfnamefont {N.}~\bibnamefont
  {Doiron-Leyraud}}, \bibinfo {author} {\bibfnamefont {J.}~\bibnamefont
  {Levallois}}, \bibinfo {author} {\bibfnamefont {R.}~\bibnamefont {Daou}},
  \bibinfo {author} {\bibfnamefont {J.-B.}\ \bibnamefont {Bonnemaison}},
  \bibinfo {author} {\bibfnamefont {N.}~\bibnamefont {Hussey}}, \bibinfo
  {author} {\bibfnamefont {L.}~\bibnamefont {Balicas}}, \bibinfo {author}
  {\bibfnamefont {B.}~\bibnamefont {Ramshaw}}, \bibinfo {author} {\bibfnamefont
  {R.}~\bibnamefont {Liang}}, \bibinfo {author} {\bibfnamefont
  {D.}~\bibnamefont {Bonn}}, \emph {et~al.},\ }\href@noop {} {\bibfield
  {journal} {\bibinfo  {journal} {Nature}\ }\textbf {\bibinfo {volume} {450}},\
  \bibinfo {pages} {533} (\bibinfo {year} {2007})}\BibitemShut {NoStop}%
\end{thebibliography}%

\end{document}